\documentclass{JHEP3}
\usepackage{amsmath}
\usepackage{epsfig}
\usepackage{cite}

\title{Aspects of power corrections in hadron-hadron collisions}

\author{Mrinal Dasgupta and Yazid Delenda\\
        School of Physics and Astronomy, University of Manchester,\\
        Oxford road, Manchester M13 9PL, U.K.\\
        \email{Mrinal.Dasgupta@manchester.ac.uk},
        \email{yazid@hep.man.ac.uk}}

\preprint{MAN/HEP/2007/14}

\abstract{The program of understanding inverse-power law corrections
to event shapes and energy flow observables in $e^{+}e^{-}$
annihilation to two jets and DIS (1+1) jets has been a significant
success of QCD phenomenology over the last decade. The important
extension of this program to similar observables in hadron
collisions is not straightforward, being obscured by both conceptual
and technical issues. In this paper we shed light on some of these
issues by providing an estimate of power corrections to the
inter-jet $E_t$ flow distribution in hadron collisions using the
techniques that were employed in the $e^{+}e^{-}$ annihilation and
DIS cases.}


\begin{document}

\section{Introduction}

The principle issue that limits the accuracy of theoretical
predictions in QCD is the presence of non-perturbative physics
responsible for the confinement of quarks and gluons. The success of
perturbative QCD, despite a lack of quantitative understanding of
the confinement process, is a major achievement that is based on
identifying infrared and collinear-safe observables \cite{SW} which
are as insensitive as possible to non-perturbative effects such as
hadronisation. In such instances the role of non-perturbative
effects is reduced to the level of corrections to the perturbative
estimates which take the form of an inverse-power law in the hard
scale $Q$, $1/Q^p$, where $p$ depends on the observable. For some
observables, such as the total cross-section in $e^{+}e^{-}$
annihilation to jets, where $p=4$, the corrections in question are
insignificant and can safely be ignored in comparisons to data. For
other observables such as event-shape variables (see ref.
\cite{DSreview1} for a review) it was noted however that the power
corrections scale as $1/Q$ and obscure the perturbative analysis
significantly.

Over the past decade, theoretical efforts essentially based on
renormalons (see ref. \cite{Beneke} and references therein) have
given a clearer picture of the origin and role of power corrections.
Within the renormalon model these corrections are shown to be
related to a factorial divergence of the perturbative QCD expansion
at high orders and the consequent error in truncating what is in
fact an asymptotic series \cite{Beneke}. This observation allows one
to estimate power corrections from a lowest-order Feynman graph,
modified to incorporate the relevant subset of higher-order terms
(renormalon bubble insertions).

From a phenomenological viewpoint, perhaps the most widely used
formulation of the renormalon model is one that uses a dispersive
representation of the running coupling \cite{Renormalon}, thereby
introducing a dispersive variable $m$ that plays the role of a fake
gluon mass \cite{BBM}. This scale is a natural trigger for power
corrections arising in the infrared regime for observables that are
otherwise dominated by a hard scale $Q$. The most appealing feature
of the dispersive approach is the hypothesis of a universal
infrared-finite coupling $\alpha_{s}$. The overall size of the power
corrections in this approach is related to moments of this universal
coupling: $\int_0^{\infty} dm^2/m^2 \alpha_{s}(m^2)(m^2/Q^2)^p$. For
most event-shape variables (for example) $p=1/2$ \cite{DSreview1},
and the relevant coupling moment or equivalently the related
quantity $\alpha_0$ has been extracted from data for different event
shapes in both $e^{+}e^{-}$ annihilation and DIS \cite{NPReview,
NPReview2}. The values of $\alpha_0$ thus obtained have generally
confirmed the universality hypothesis to within the expected
uncertainties due to missing higher-order corrections
\cite{DSreview1}. These studies have not only enabled successful
studies of several observables, but also lent credence to the notion
that the QCD coupling may be a meaningful (finite and universal)
concept all the way down to the smallest energy scales, and have
thus renewed hope of a better understanding of the confinement
domain from first principles of QCD.

The success of renormalon-inspired studies has thus far been limited
to observables which involve just two hard partons at Born level
such as event shapes in $e^{+}e^{-} \rightarrow $ two jets and DIS
($1+1$) jets. Some of the most interesting QCD observables however
do not fall into the above class. Examples include three-jet event
shapes in $e^+e^-$ annihilation and DIS \cite{Power3jetDokshitzer1,
Power3jetDokshitzer2, Power3jetDokshitzer3, Power3jetBanfi}, dijet
event shapes in hadron collisions \cite{BSZ} and single-jet
inclusive cross-sections at hadron colliders \cite{Power4jet}. These
observables contain more than two hard partons at Born level and
additionally pertain to processes that have gluons at Born level
where one may question the extension of the dispersive approach,
which (strictly speaking) was formulated for observables involving
only quarks at the Born level \cite{Renormalon}. The program of
understanding power corrections has not been put to test in such
situations which will be of importance at the LHC, for instance,
although some progress is being made on the phenomenological side
for three-jet event shapes \cite{Power3jetBanfi}. In fact for the
case of observables involving four hard partons at the Born level
there are as yet no theoretical predictions for power corrections.

Applying the renormalon model, in many such cases, reduces (as for
$e^{+}e^{-}$ annihilation to jets or DIS event shapes) to an
analysis of soft gluon radiation with transverse momenta $k_t \sim
\Lambda_{\mathrm{QCD}}$, which are associated with a universal
infrared-finite coupling. Given the rich structure (colour and
geometry-dependence) of soft gluon radiation for processes like
dijet production in hadron collisions, it is indeed enticing to see
how a perturbative structure may influence predictions for power
corrections as predicted by the renormalon model. Signs of
perturbatively calculable colour structure in the non-perturbative
power-behaved component would strongly suggest that the
renormalon-inspired picture of these corrections is an extension of
soft gluon radiation, with a modified but universal coupling,
thereby conclusively establishing the model.

In this paper we provide a calculation of the power correction to
the transverse energy ($E_t$) flow away from jets accompanying hard
dijet production in hadron collisions. A variant of this observable
(the ``pedestal height'') was suggested several years ago as a means
of separating perturbative bremsstrahlung from the contribution of
the soft underlying event in hadron collisions \cite{MW}. To date
however a satisfactory understanding of the away-from-jet energy
flow has proved elusive. In the region of small $E_t$ the
distribution $d\sigma/dE_t$ contains large logarithms of the form
$\alpha_s^n \ln^{n-1}(P_t/E_t)/E_t$, where $P_t$ is the hard scale
(jet transverse momenta) of the problem. These logarithms can only
be resummed in the large $N_c$ limit \cite{DassalNG1,DassalNG2,BMS}
which limits the accuracy of perturbative estimates that can be made
in this case alongside the lack of any estimates of the NLL
contributions. One may also expect power corrections of the form
$1/P_t$ to the resummed distribution as for the case of event-shape
distributions in $e^{+}e^{-}$ annihilation and DIS. Once again there
is no estimate for these power corrections and it is this that we
aim to provide here. Combining the power corrections we compute with
resummed $E_t$ flow distributions gives a more complete theoretical
account which should facilitate comparisons to experiment.

While at hadron colliders the ever-present soft underlying event may
obstruct clean studies of power-behaved corrections arising from the
bremsstrahlung component of the $E_t$ flow, our calculations here
are also easily adapted to the case of rapidity gaps in dijet
photoproduction at HERA, and are in principle more readily tested in
that environment. Moreover there are theoretical ideas concerning
observables such as inclusive jet cross-sections at hadron
colliders, where it may be possible to disentangle the underlying
event from the non-perturbative physics in the bremsstrahlung
component, due to a singular $1/R$ ($R$ being a jet radius
parameter) behaviour of the latter \cite{Power4jet}. A full estimate
of this piece however involves a calculation similar to the one we
introduce here and thus the present calculations can be used as a
guide in moving towards better estimates of the inclusive jet
cross-sections as well. Another related class of observables, to
which the techniques we use here are directly applicable, is the
important case of event shapes at hadron colliders \cite{BSZ}, for
which resummed perturbative estimates already exist and identical
considerations to those of this paper will be required when dealing
with the issue of power corrections. While other problems such as
lack of knowledge about next-to-leading logarithms also need to be
addressed, the calculation of the power correction is ultimately an
important ingredient which, as we stressed above, also serves as a
case study for hadron-hadron observables more generally.

This paper is organised as follows. In the following section we
define the observable more precisely and review the perturbative
result in Mellin space conjugate to $E_t$, which involves colour
matrices in the resummed anomalous dimensions
\cite{AnomalousMatrices1, AnomalousMatrices2, AnomalousMatrices3,
AnomalousMatrices4}. We then compute the power correction to each of
the matrix elements of the anomalous dimension by considering the
appropriate combination of dipoles involved in that matrix element.
The calculation of the power correction in each dipole term is
performed using the appropriate scale of the running coupling (the
invariant transverse momentum of the dipole, $k_{\perp}$
\cite{Power3jetDokshitzer1, Power3jetDokshitzer2,
Power3jetDokshitzer3, Banfi, CG, SterDix}). The final step is to
take the inverse Mellin transform of the result including
non-perturbative corrections, for which it proves convenient to
diagonalise the power-corrected anomalous dimension matrices.We find
that the power correction to the $E_t$ distribution is not a simple
shift of the resummed distribution by a fixed amount proportional to
$1/P_t$ as is the case for event shapes in $e^{+}e^{-}\rightarrow$
two jets and DIS (to a good approximation).

\section{Resummed perturbative result}

Here we outline the resummed result for the $E_t$ flow distribution
accompanying hard dijet production in hadronic collisions, which was
originally computed in ref.~\cite{BKS}, specialising for simplicity
to the case of a slice in rapidity of width $\Delta \eta$, which we
take to be centred at $\eta=0$ (the rapidity $\eta$ and transverse
energy $E_t$ are both defined with respect to the beam direction).
We also work, again purely for the sake of convenience, with fixed
jet transverse momenta $P_t$ and rapidities but can easily
generalise to other differential distributions of the hard dijet
system. The observable we compute is then the integrated
distribution:
\begin{equation}
\Sigma(P_t,E_t) = \frac{1}{\sigma_0} \int_0^{E_t}
\frac{d\sigma}{dE_t'} dE_t', \qquad E_t = \sum_{i\in\Omega} E_{t,i},
\end{equation}
where $\sigma_0$ is the Born cross-section for dijet production (at
\emph{fixed jet transverse momenta and rapidities}) in hadron
collisions, $\Omega$ denotes the rapidity slice and $E_t$ is
obtained by summing over all objects (partons/jets) in the gap.

Since we are examining the accompanying $E_t$ flow distribution, we
are dealing typically with $E_t \ll P_t$ and hence there arises the
need to resum large logarithms of the form $\alpha_s^n \ln^{m}
(P_t/E_t)$, with $m \leq n$. The resummed prediction for
$\Sigma(P_t,E_t)$ up to single-logarithmic accuracy (resumming all
the terms $\alpha_s^n \ln^n (P_t/E_t)$ in the logarithm of the
integrated distribution), can be thought of as comprising two
distinct pieces with different dynamical origins: ``global'' and
``non-global''\footnote{We ignore, for now, a possible additional
complication afflicting non-global observables in hadronic dijet
production -- super-leading logarithms \cite{FKS} that may appear at
fourth-loop level. The status of these is as yet unclear.}
\cite{DassalNG1, DassalNG2}.

The first piece (the so-called ``global'' component) is a result of
considering multiple soft emissions, both real and virtual, which
are attached just to the primary or Born hard partons. Due to
infrared-safety of the observable real and virtual emissions cancel
below the scale $E_t$, while real emissions above this scale are
vetoed. Thus the resummed result for this piece is just the
summation of virtual graphs above the scale $E_t$, attached to the
primary hard partons \cite{AnomalousMatrices1, AnomalousMatrices2,
AnomalousMatrices3, AnomalousMatrices4}. In actual fact the
factorisation of real emissions and the consequent cancellation with
virtual ones takes place in Mellin space conjugate to $E_t$. This
complication can be ignored for the leading single-logarithmic
terms, but to analyse the impact of power corrections on the
resummed distribution we need to compute the result in Mellin space
and then invert the transform to $E_t$ space. To be more precise the
perturbative resummed result (considering just the above described
global term for now) reads (see for instance ref. \cite{BMS}):
\begin{equation} \label{eq:anorm}
\Sigma(P_t,E_t) = \int \frac{d \nu}{2 \pi i \nu} e^{\nu E_t}
\mathcal{R}(\nu), \,
\end{equation}
where $\nu$ is the Mellin variable conjugate to $E_t$ and the
integration contour is taken, in the usual manner, parallel to the
imaginary axis and to the right of all singularities of the
integrand. We have:
\begin{equation}
\label{eq:global} \mathcal{R}(\nu) = \mathrm{Tr}\left(\mathbf{H}
e^{-\mathbf{\Gamma}^{\dagger}(\nu)} \mathbf{S}
e^{-\mathbf{\Gamma}(\nu)}\right)/\Sigma_0.
\end{equation}
In the above $\mathbf{\Gamma}$ is essentially an ``anomalous
dimension matrix'' and $\mathbf{H}$ and $\mathbf{S}$ are the
``hard'' and ``soft'' matrices. The matrix elements $H_{ij}$
represent the product of the Born amplitude in colour channel $i$
and its complex conjugate in colour channel $j$, and the matrix
$\mathbf{S}$ represents the normalisation arising from the colour
algebra. The squared matrix element for the Born scattering in this
notation is just $\Sigma_0=\mathrm{Tr}(\mathbf{H}\mathbf{S})$. We
shall return to the detailed structure of $\mathbf{\Gamma}$ in the
next section but it is important to point out here that all the
matrices above depend on the partonic $2 \to 2$ sub-process
considered, e.g. they differ for say $q \bar{q} \to q \bar{q}$ and
$gg \to gg$ sub-processes. Their form also depends on one's choice
of colour basis. For examples the reader is pointed to the original
references \cite{AnomalousMatrices1, AnomalousMatrices2,
AnomalousMatrices3, AnomalousMatrices4}. We also point out that
since we deal exclusively with soft radiation and differential
distributions in the jet rapidities and transverse momenta, the
parton distribution functions simply factor from the resummed result
and cancel against those in the normalisation factor $\sigma_0$.

We now turn to the second piece of resummed distribution: the
``non-global'' contribution. The above result, resumming essentially
virtual corrections above the veto scale $E_t$ which dress the hard
scattering, is not the complete description at single-logarithmic
accuracy \cite{DassalNG1,DassalNG2}. An additional ``non-global''
piece $\mathcal{S}(P_t,E_t)$ arises (starting at the two-gluon
emission level, $\mathcal{O}(\alpha_s^2)$) and is also
single-logarithmic. The dynamical origin of this piece is multiple
soft energy-ordered radiation with an arbitrary complex geometry
(``hedgehog'' configurations of soft gluons, as opposed to emissions
strongly ordered in angle). It has thus far been possible to treat
this term only in the large $N_c$ approximation, which limits the
accuracy of perturbative calculations in the present instance.

However it has been pointed out and clarified in a series of papers
\cite{AS1, BD05, ktc, AS2} that the role of the non-global component
can be significantly reduced by defining the observable in terms of
soft jets rather than individual hadrons in the gap. This can be
achieved by running a clustering algorithm on the final states such
that all objects are included in jets. The clustering procedure
(with a large cluster radius $R=1$) was shown to virtually eliminate
the non global component while giving rise to additional global
terms \cite{ktc} that were at most modest corrections to the pure
virtual dressing as represented here by eq.~\eqref{eq:global}.
Moreover the power corrections associated to the non-global
component of the result would start at a higher-order in $\alpha_s$
(albeit potentially accompanied by logarithms of $P_t/E_t$) which we
ignore. This was also the procedure employed for the case of
non-global DIS event shapes \cite{DSreview1}, where power
corrections were computed in the exponentiated single-gluon piece of
the resummed distribution, which was a phenomenological success. For
all these reasons we shall choose to concentrate  for the rest of
this paper on the global component (eq.~\eqref{eq:global}), ignoring
the non-global component.

In the following section we explicate the structure of
$\mathbf{\Gamma}$ in terms of the various hard colour dipoles from
which one considers soft gluons to be emitted according to the usual
antenna pattern. Non-perturbative power corrections are then
computed on a dipole-by-dipole basis, adapting the procedure for a
$q \bar{q}$ dipole developed for the case of $e^{+}e^{-}$
annihilation to two jets. Having obtained the power corrections to
$\mathbf{\Gamma}(\nu)$ we then invert the Mellin transform to
examine the result for the $E_t$ distribution.

\section{The anomalous dimension and power corrections}

We first write down the structure of the resummed anomalous
dimension matrix $\mathbf{\Gamma}(\nu)$ and then note that it
contains an integral over the running coupling which is formally
divergent. Making the ansatz of a universal infrared-finite coupling
cures this divergence and introduces calculable power corrections to
the perturbative anomalous dimensions. In what follows and for the
rest of this paper we specialise to the case of the sub-process $q
\bar{q} \to q \bar{q}$ since identical considerations are involved
for all other sub-processes. Full results combining all
sub-processes will be made available in forthcoming work.

For the sub-process $q\left (p_1,r_1 \right)+\bar{q}\left (p_2,r_2
\right) \to q \left(p_3,r_3 \right)+\bar{q}\left(p_4,r_4 \right)$,
where $p_i$ and $r_i$ are respectively four-momenta and colour
indices, we can choose to work in the $t$-channel singlet-octet
colour basis:
\begin{equation}
c_{1} = \delta_{r_1 r_3} \delta_{r_2 r_4}, \qquad c_{2} =
\frac{1}{2} \left (\delta_{r_3 r_4}\delta_{r_1 r_2}-\frac{1}{N_c}
\delta_{r_1 r_3} \delta_{r_2 r_4} \right ).
\end{equation}
For this basis we have:
\begin{equation}
\mathbf{S}=\left(
\begin{array}{ccc} N_c^2 & 0\\
0 & \frac{N_c^2-1}{4}
\end{array}\right),
\end{equation}
and the anomalous dimension matrix $\mathbf{\Gamma}$ is:
\begin{equation}
\label{eq:anom} \mathbf{\Gamma} =\left(
\begin{array}{ccc} C_F T& \frac{C_F}{2N_c}(S-U)\\
S-U & C_F S-\frac{1}{2N_c}(T-2U+S)
\end{array}\right).
\end{equation}
In the above $S$, $T$ and $U$ are combinations of dipole
contributions with each contribution given by the corresponding
dipole antenna. Thus one has\footnote{Here our use of the $S$, $T$
and $U$ labels differs slightly from that of other references, e.g.
\cite{AnomalousMatrices1, AnomalousMatrices2, AnomalousMatrices3,
AnomalousMatrices4}.}:
\begin{eqnarray}
S &= \tilde{w}_{12}+\tilde{w}_{34}, \\
T &= \tilde{w}_{13}+\tilde{w}_{24}, \\
U & = \tilde{w}_{23}+\tilde{w}_{14},
\end{eqnarray}
where each dipole contribution $\tilde{w}_{ij}$ reads:
\begin{equation}
\tilde{w}_{ij} = \int \frac{d^3 \vec{k}}{2\omega(2\pi)^3} g_s^2
\frac{\left(p_i.p_j\right)}{\left(p_i.k\right)\left(p_j.k \right)}
u(k_t,\nu).
\end{equation}
In the above result we have integrated the soft gluon emission
probability, given by the dipole antenna pattern, over the gluon
phase-space (with $\vec{k}$ and $\omega$ being respectively the
three-momentum and energy of the gluon, as usual) with a ``source''
function $u(k_t, \nu)$ and $g_s^2 =4 \pi \alpha_s$. The source is a
result of factorising the real soft emission phase-space in Mellin
space (see for instance ref.~\cite{BMS}) and accounting additionally
for virtual corrections. It reads:
\begin{equation}
u(k_t,\nu) = \left(1-e^{-\nu k_t}\right ),
\end{equation}
if the emission is in $\Omega$ and zero elsewhere. The source thus
represents the impact of real-virtual contributions which completely
cancel, to our accuracy, for emissions outside $\Omega$. Now
introducing the variables $\eta$ and $\phi$, respectively the
rapidity and azimuth of the emission with respect to the beam
direction, one can write:
\begin{equation}
\label{eq:dip} \tilde{w}_{ij} = \int \frac{d k_t}{k_t}
\frac{\alpha_s(k_{\perp,i,j}^2)}{2 \pi} d\eta \frac{d\phi}{2\pi}
\left(1-e^{-\nu k_t}\right) f_{ij}(\eta,\phi),
\end{equation}
with $f_{ij} (\eta, \phi)$ being the functional dependence on
rapidity and azimuth that arises from the dipole antenna patterns
and which we shall use below (see appendix). We note that, to our
accuracy, the correct argument of the running coupling for emission
from a dipole (\cite{Power3jetDokshitzer1, Power3jetDokshitzer2,
Power3jetDokshitzer3, Banfi, CG,SterDix}) is $k_{\perp,ij}^2=
2(p_i.k)(p_j.k)/(p_i.p_j)$, which is the transverse momentum of the
gluon $k$ with respect to the dipole axis in the dipole rest frame.
This must be distinguished from $k_t$, the transverse momentum of
the gluon with respect to the beam direction, which is the quantity
that directly enters the observable definition. In fact we have, in
terms of the functions $f_{ij}$ introduced above, $k_{\perp,ij} =
k_t \sqrt{2/f_{ij}}$.

\subsection{Power corrections dipole-by-dipole}

Now we proceed to an extraction of the leading power-behaved
contribution. In order to do this we first note that the integral
over $k_t$ in eq. \eqref{eq:dip}, which can be rewritten as one over
the related variable $k_{\perp}$, is divergent if one uses the usual
perturbative definition of $\alpha_s$, due to the divergence of the
perturbative running coupling at $k_{\perp} =
\Lambda_{\mathrm{QCD}}$. In order to isolate and cure this
pathological behaviour we assume an infrared-finite coupling
($\alpha_{s}$) and change variable of integration from $k_t$ to
$k_{\perp}$ in eq. \eqref{eq:dip}. We then follow the method of ref.
\cite{PowerEventShapes} to write $\alpha_{s}(k_{\perp}^2) =
\alpha_{s, \mathrm{PT}}(k_{\perp}^2)+ \delta \alpha_{s,\mathrm{NP}}
(k_{\perp}^2)$, where PT and NP stand for perturbative and
non-perturbative respectively. In doing so we have assumed that the
actual coupling $\alpha_{s}$ is in fact finite even at arbitrarily
small $k_{\perp}$, and can be split into the usual perturbative
component $\alpha_{s,\mathrm{PT}}$ and a modification $\delta
\alpha_{s,\mathrm{NP}}$ which is due to non-perturbative effects.
Both the perturbative and non-perturbative components separately
diverge, but the divergences cancel in their sum due to the assumed
finiteness of the physical coupling $\alpha_{s}$. Moreover, since we
do not wish to modify the perturbative results at large scales, the
non-perturbative physics as represented by the modification $\delta
\alpha_{s,\mathrm{NP}}$ must vanish above some infrared ``matching''
scale $\mu_I$. Effectively the addition of the $\delta
\alpha_{s,\mathrm{NP}}$ term represents removal of the badly-behaved
perturbative contribution below $\mu_I$ and its replacement with the
well-behaved integral over the infrared-finite physical coupling
$\alpha_s$.

Thus for the observable itself one has from dipole ($ij$):
\begin{multline}
\label{eq:ptnp} \int_{k \in \Omega} d\eta \frac{d\phi}{2\pi} \frac{d
k_{\perp,i,j}}{k_{\perp,i,j}} \frac{\alpha_{s,\mathrm{PT}}
(k_{\perp,i,j}^2)}{2\pi} \left(1-e^{-\nu k_t}\right)
f_{ij}(\eta,\phi)+\\+ \int_{k \in \Omega} d\eta \frac{d\phi}{2\pi}
\frac{dk_{\perp,i,j}}{k_{\perp,i,j}} \frac{\delta
\alpha_{s,\mathrm{NP}}(k_{\perp,i,j}^2)}{2\pi} \left(1-e^{-\nu
k_t}\right)f_{ij}(\eta,\phi).
\end{multline}
The integral involving the perturbative coupling represents the
usual perturbative contribution from dipole ($ij$). The leading
logarithmic perturbative contribution arises from the region where
one can make the approximation:
\begin{equation}
\label{eq:subs} \left(1-e^{-\nu k_t} \right) \approx
\theta\left(k_t-\frac{1}{\nu}\right).
\end{equation}
The perturbative results are reported at length in ref.~\cite{BKS}.
In what follows we shall consider in more detail the
non-perturbative contribution from the integral involving $\delta
\alpha_{s,\mathrm{NP}}$.

In order to evaluate the non-perturbative contribution we first
consider that the leading such term arises from the region $\mu_I
\ll 1/\nu$, which translates to a requirement on $E_t$ to be above a
few GeV. In this region one can expand the exponential in an exactly
analogous way as for event-shape distributions
\cite{PowerEventShapes}. The leading term is given by the first term
in the expansion: $1-\exp(-\nu k_t)\approx \nu k_t$, and this
corresponds to a linear $1/P_t$ power correction. We ignore
quadratic and higher power corrections that would scale as $1/P_t^2$
and beyond, once again following the case of event-shape variables.
We also note that in the shape function approach \cite{KS94,KorSter,
KorSter2}, where one may study non-perturbative effects even into
the region $E_t \sim \Lambda_{\mathrm{QCD}}$, higher powers of $\nu$
need also to be retained. Working with just the leading term gives
us the non-perturbative correction from the ($ij$) dipole which can
then be written as:
\begin{equation}
\label{eq:dipshift} \tilde{w}_{ij} = \tilde{w}_{ij}^{\mathrm{PT}}
+\nu {\mathcal{P}} C_{ij}.
\end{equation}
Here the non-perturbative quantity ${\mathcal{P}}$ is the first
moment of the coupling modification $\delta \alpha_{s,\mathrm{NP}}$:
\begin{equation}
\mathcal{P} = \int_0^{\mu_I}\frac{dk_{\perp}}{k_{\perp}} k_{\perp}
\frac{\delta \alpha_{s,\mathrm{NP}}(k_{\perp}^2)}{2\pi},
\end{equation}
which also enters $1/Q$ ($Q$ being the hard scale) power corrections
to event shapes and can be related to the parameter $\alpha_0$,
extracted from fits to event shape data, as:
\begin{equation}
\mathcal{P}=\frac{\mu_I}{2\pi}\left(\alpha_0(\mu_I)
-\alpha_s(P_t^2)-\frac{\beta_0}{2\pi}\left(\ln\frac{P_t}{\mu_I}+\frac{K}
{\beta_0}+1\right)\alpha_s^2(P_t^2)+\mathcal{O}(\alpha_s^3)\right),
\end{equation}
where $\alpha_0(\mu_I)=1/\mu_I\int_0^{\mu_I}dk_{\perp}
\,\alpha_{s}(k_{\perp}^2)$, $\beta_0$ is the first coefficient of
the QCD beta function and:
\begin{equation}
K = C_A \left(\frac{67}{18}-\frac{\pi^2}{6}\right)-\frac{5}{9}n_f.
\end{equation}

The coefficients $C_{ij}$ represent the integral over
directions:
\begin{equation}
\label{eq:int} C_{ij} = \int_{k \in \Omega} d\eta \frac{d\phi}{2\pi}
\frac{1}{\sqrt{2}}f_{ij}^{3/2},
\end{equation}
where $f_{ij}$ arises from the dipole antenna pattern as indicated
in eq. $\eqref{eq:dip}$, and a further factor proportional to
$\sqrt{f_{ij}}$ comes from rewriting $k_t$ in terms of $k_{\perp}$
as we stated before. The explicit form of the $f_{ij}$ functions is
reported in the appendix. Performing the integrals over $\eta$ and
$\phi$ in eq.~\eqref{eq:int} yields the coefficients $C_{ij}$ that
correspond to the non-perturbative contribution to $\tilde{w}_{ij}$
in eq.~\eqref{eq:dipshift}, which we do not explicitly display for
economy of presentation.

Having computed the power corrections proportional to $\nu$ for each
dipole, we can include these corrections to the anomalous dimension
matrix in eq.~\eqref{eq:anom}, which can then be written as:
\begin{equation}
\mathbf{\Gamma} = \tau(\nu) \mathbf{\Gamma}_{\mathrm{PT}}+\nu
\mathcal{P} \mathbf{\Gamma}_{\mathrm{NP}},
\end{equation}
where the non-perturbative contribution
$\mathbf{\Gamma}_{\mathrm{NP}}$ is built up by combining the dipole
contributions $C_{ij}$ as in the perturbative case. In the case of
the perturbative term $\mathbf{\Gamma}_{\mathrm{PT}}$ we explicitly
extracted the integral over the transverse momentum of the coupling:
\begin{equation}
\tau(\nu) = \int_{1/\nu}^{P_t} \frac{dk_t}{k_t}
\frac{\alpha_{s,\mathrm{PT}}(k_t^2)}{2\pi},
\end{equation}
which arises by making the substitution \eqref{eq:subs} in the first
term of \eqref{eq:ptnp}. Then the matrix $\mathbf{\Gamma
}_{\mathrm{PT}}$ is the usual perturbative anomalous dimension
containing integrals over gluon directions inside the
region\footnote{These integrals are similar to those which yield the
$C_{ij}$ except that the functions $f_{ij}$ are involved rather than
$f_{ij}^{3/2}/\sqrt{2}$.} $\Omega$. In the following section we
shall consider the evaluation of the inverse Mellin transform to
take our results from $\nu$ space to $E_t$ space.

\section{Power corrections in the $E_t$ cross-section}

After accommodating the leading power corrections (those expected to
give rise to $1/P_t$ effects) eq.~\eqref{eq:anorm} assumes the
explicit form:
\begin{multline}
\Sigma(P_t,E_t) = \int \frac{d\nu}{2\pi i \nu} e^{\nu E_t} \times\\
\times\mathrm{Tr}\left[\mathbf{H} \exp\left(-\tau
\mathbf{\Gamma}_{\mathrm{PT}}^{\dag} - \nu \mathcal{P}
\mathbf{\Gamma}_{\mathrm{NP}}^{\dag}\right) \mathbf{S}
\exp\left(-\tau \mathbf{\Gamma}_{\mathrm{PT}}-\nu \mathcal{P}
\mathbf{\Gamma}_{\mathrm{NP}}\right) \right]/\Sigma_0.
\end{multline}

In order to invert the Mellin transform (perform the $\nu$ integral
above) it is simplest to diagonalise the matrix $\tau \mathbf{\Gamma
}_{\mathrm{PT}}+\nu \mathcal{P} \mathbf{\Gamma}_{\mathrm{NP}}$. In
the basis in which the matrix $\mathbf{\Gamma}$ is diagonal the
matrices $\mathbf{H}$ and $\mathbf{S}$ become
$\widetilde{\mathbf{H}} = \mathbf{R}^{-1} \mathbf{H}
\mathbf{R}^{-1\dag}$ and $\widetilde{\mathbf{S}} =
\mathbf{R}^{\dag}\mathbf{S}\mathbf{R}$, where $\mathbf{R}$ is a
matrix which contains the eigenvectors of $\mathbf{\Gamma}$ as
column entries (see also \cite{FKS2}). After diagonalisation we can
write the result for $\Sigma(P_t,E_t)$ in terms of components as:
\begin{equation}
\Sigma(P_t,E_t) = \int \frac{d\nu}{2 \pi i \nu} e^{\nu E_t}
e^{-\lambda_{i}} \delta_{ij} \widetilde{H}_{jk}e^{-\lambda^*_{k}}
\delta_{kl}\widetilde{S}_{li}/\Sigma_0,
\end{equation}
where $\lambda_i$ are the eigenvalues of the matrix
$\mathbf{\Gamma}$. For the case of the $q\bar{q} \to q \bar{q}$
sub-process, which we use as an example, $\mathbf{\Gamma}$,
$\widetilde{\mathbf{H}}$ and $\widetilde{\mathbf{S}}$ are $2\times2$
matrices and the above result can be written explicitly in terms of
the elements of the various matrices as:
\begin{multline}
\label{eq:trace}\Sigma(P_t,E_t)= \int \frac{d\nu}{2 \pi i \nu}
e^{\nu E_t}\frac{1}{\Sigma_0} \times\\ \times \left(
\widetilde{H}_{11}\widetilde{S}_{11}e^{-(\lambda_1+\lambda_1^*)}
+\widetilde{H}_{12}\widetilde{S}_{21}e^{-(\lambda_1+\lambda_2^{*})}
+\widetilde{H}_{21}\widetilde{S}_{12}e^{-(\lambda_1^{*}+\lambda_2)}
+\widetilde{H}_{22}\widetilde{S}_{22}
e^{-(\lambda_2+\lambda_2^*)}\right),
\end{multline}
where the above result contains both perturbative and
non-perturbative contributions. To separate these we note that the
eigenvalues can be expanded so as to retain only the first-order in
$\nu$ correction to the perturbative value, which depends
logarithmically on $\nu$:
\begin{equation}
\lambda_i = \tau(\nu) \lambda_i^{\mathrm{PT}} +  \nu \mathcal{P}
\lambda_i^{\mathrm{NP}}+\mathcal{O}(\nu^2).
\end{equation}
We emphasise here that while $\lambda^{\mathrm{PT}}_i$ are simply
the eigenvalues of $\mathbf{\Gamma}_{\mathrm{PT}}$,
$\lambda^{\mathrm{NP}}_i$ are \emph{not} the eigenvalues of
$\mathbf{\Gamma}_{\mathrm{NP}}$. Instead they are coefficients of
the $\mathcal{O}(\nu)$ component of the expansion of the eigenvalues
of $\tau \mathbf{\Gamma}_{ \mathrm{PT}}+ \nu\mathcal{P}
\mathbf{\Gamma}_{\mathrm{NP}}$ and they depend on the components of
both $\mathbf{\Gamma}_{\mathrm{NP}}$ and $\mathbf{\Gamma
}_{\mathrm{PT}}$.

The matrices $\widetilde{\mathbf{H}}$ and $\widetilde{\mathbf{S}}$
also differ from their pure perturbative forms by corrections which
depend on $\nu$. Let us introduce the notation
$D_{ij}=\widetilde{H}_{ij}\widetilde{S}_{ji}$ (summation \emph{not}
implied). We expand the elements $D_{ij}$ to first order in $\nu$
and write the result as:
\begin{equation}
\label{eq:pt} \Sigma(P_t,E_t) =\sum_{i,j} \int \frac{d\nu}{2 \pi i
\nu} e^{\nu\left(E_t-\mathcal{P}\left[\lambda_{i}^{\mathrm{NP}}+
\lambda_j^{*\mathrm{NP}}\right]\right)} \left(D_{ij}^{\mathrm{PT}}+
\frac{\nu}{\tau(\nu)}\mathcal{P} D_{ij}^{\mathrm{NP}}\right)
e^{-\tau(\nu)(\lambda_i^{\mathrm{PT}}+\lambda_j^{*\mathrm{PT}})}
/\Sigma_0,
\end{equation}
where the sum runs over the components of the matrix $D_{ij}$ and we
ignored $\mathcal{O}(\nu^2)$ terms in the expansion of $D_{ij}(\nu
)$, as they would contribute to corrections that scale as $1/E_t^2$,
which are beyond our control with the present method. We shall see
that the terms we ignore are expected to be numerically of no
significance.

We first write down the pure perturbative result $\Sigma^{\mathrm{PT
}} (P_t,E_t)$ obtained by ignoring all non-perturbative (NP)
components. To perform the $\nu$ integral we use the standard method
of expanding $\tau(\nu)$ about the saddle point at $\nu =1/E_t$,
\begin{equation}
\tau(\nu) = \tau(1/E_t)+\ln(\nu P_t) \left(\frac{\partial\,
\tau(\nu)}{\partial \ln (\nu P_t)}\right)_{\nu=1/E_t}+\cdots.
\end{equation}
Performing the integral over $\nu$ and noting that the logarithmic
derivatives of $\tau(\nu)$ are related to subleading terms of
relative order $\alpha_s$ to the leading single logarithms, we
arrive at the leading-logarithmic resummed result:
\begin{equation}\label{eq:temp10}
\Sigma^{\mathrm{PT}} (P_t,E_t) \approx \sum_{i,j}
D_{ij}^{\mathrm{PT}} e^{-\left(\lambda_i^{\mathrm{PT}}+
\lambda_j^{*\mathrm{PT}}\right) \tau(E_t)}/\Sigma_0,
\end{equation}
where the effect of the $\nu$ integration amounts to merely
replacing $\nu$ by $1/E_t$. Now redoing the $\nu$ integral including
the non-perturbative correction to the eigenvalues (still ignoring
the NP corrections to $D_{ij}$) we find by examining
eq.~\eqref{eq:pt} that the impact of the non-perturbative term
amounts simply to a shift of the perturbative result in each of the
terms in the sum in eq. \eqref{eq:temp10}:
\begin{equation}
E_t \rightarrow  E_t- \mathcal{P}
\left(\lambda_i^{\mathrm{NP}}+\lambda_j^{*\mathrm{NP}}\right).
\end{equation}
Looking at the distribution in $E_t$, with $E_t$ being measured in
units of the hard scale $P_t$, amounts to a $1/P_t$ non-perturbative
shift in each term in the sum above, as is the case for two-jet
event-shape variables \cite{KorSter, PowerEventShapes}. However in
contrast to the case of event shapes it should be clear that the
overall impact of the power correction is not simply a shift of the
perturbative distribution by a fixed amount since each term in the
sum on the right hand side of eq.~\eqref{eq:temp10} receives its own
characteristic shift depending on the sum of the eigenvalues
$\lambda_i^{\mathrm{NP}}+ \lambda_j^{*\mathrm{NP}}$ entering the
term in question.

We have still not accounted for the non-perturbative contribution to
the colour basis as contained in the $D_{ij}^{\mathrm{NP}}$ terms.
To evaluate these one performs the contour integral in question
which yields a power correction of the form $\mathcal{P}/E_t$, which
is related to the fact that these pieces are proportional to $\nu$.
Computing the full result for this piece using precisely the same
expansion about $\nu = 1/E_t$ as before and discarding perturbative
subleading terms beyond single-logarithmic accuracy, we arrive at
our final result for $\Sigma(P_t,E_t)$:
\begin{equation}
\label{eq:result_Eflow} \Sigma(P_t,E_t) = \sum_{i,j}
e^{-\Delta_{ij}^{\mathrm{PT}} \tau
\left(E_t-\Delta_{ij}^{\mathrm{NP}}\right)}
\left[D_{ij}^{\mathrm{PT}}+D_{ij}^{\mathrm{NP}}
\frac{\mathcal{P}}{E_t}\frac{1}{\tau(E_t)}G_{ij}
\left(\alpha_s,\frac{P_t} {E_t}\right) \right]/\Sigma_0,
\end{equation}
where $\Delta_{ij}^{\mathrm{NP}}=\mathcal{P}
(\lambda_i^{\mathrm{NP}}+\lambda_j^{*\mathrm{NP}})$ and
$\Delta_{ij}^{\mathrm{PT}}=\lambda_i^{\mathrm{PT}}
+\lambda_j^{*\mathrm{PT}}$, and the function $G_{ij}$ is
approximately a constant of order $\alpha_s(P_t)$, varying very
slowly with $E_t$ over the range of $E_t$ we consider here. It is a
function of the logarithmic derivative of the single-log resummed
perturbative result and hence scales as $\alpha_s$,
\begin{equation}
G_{ij} \equiv \frac{\partial}{\partial \ln(P_t/E_t)} \left
(\Delta_{ij}^{\mathrm{PT}} \tau(E_t)+\ln(\tau \left (E_t \right)
\right) \approx \alpha_s(P_t).
\end{equation}
We also note the presence of $1/\tau(E_t)$ accompanying the $1/E_t$
dependence above, which is a reflection of the fact that the
correction terms (involving $D_{ij}^{\mathrm{NP}}$) go as
$\nu/\tau$.

\section{Results and conclusions}

In this section, for completeness, we illustrate the impact of
non-perturbative power corrections on the energy flow distribution
we discussed above.

\FIGURE[ht]{\epsfig{file=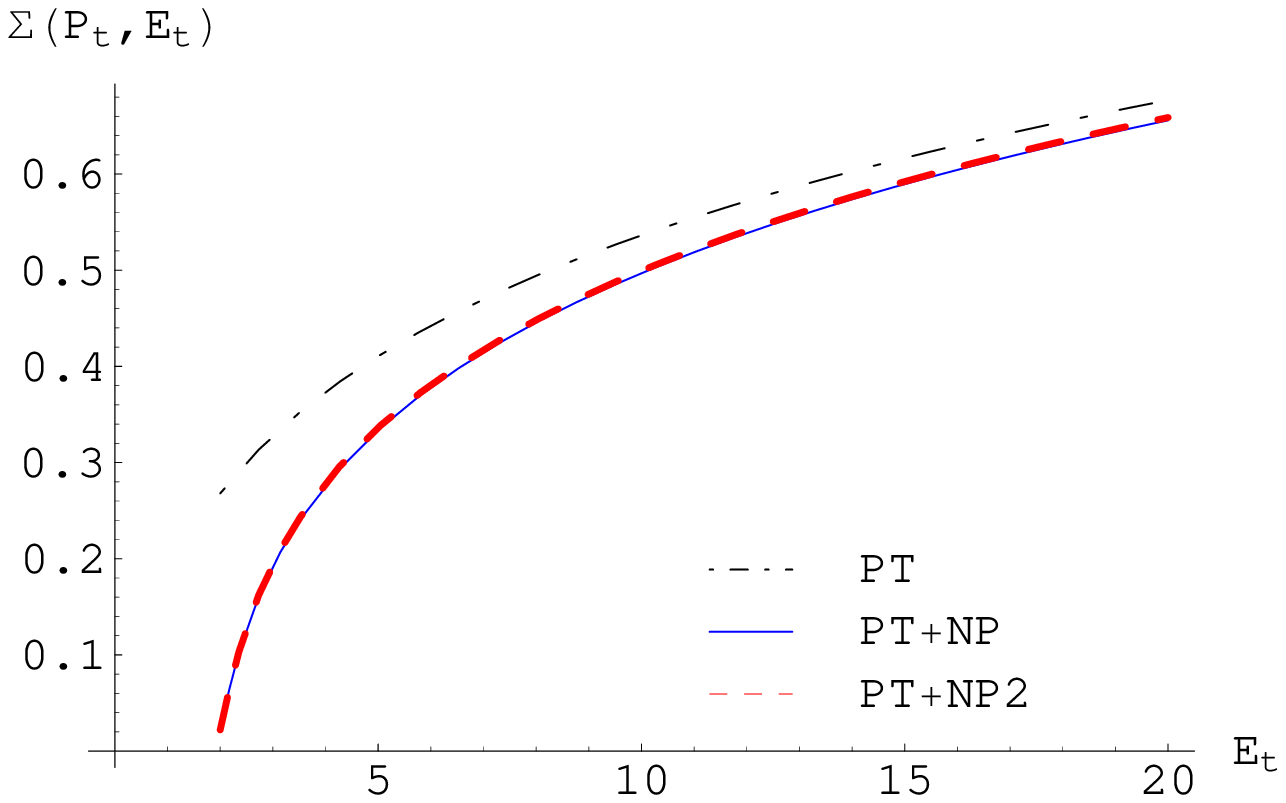,width=0.7\textwidth} \caption{Power
corrections to the energy flow distribution. PT stands for the pure
perturbative result $\Sigma^{\mathrm{PT}}(P_t,E_t)$ as presented in
eq. \eqref{eq:temp10}, PT+NP stands for the result including the
non-perturbative correction and ignoring the terms
$D_{ij}^{\mathrm{NP}}$, while PT+NP2 stands for the result presented
in eq. \eqref{eq:result_Eflow}.}\label{fig:power}}

In fig.~\ref{fig:power} we show the result for $\Sigma(P_t,E_t)$. We
present the pure perturbative result (eq. \eqref{eq:temp10}), the
result including non-perturbative corrections without the
$D_{ij}^{\mathrm{NP}}$ component (PT+NP) and the result presented in
eq.~\eqref{eq:result_Eflow} (PT+NP2).

The results above were obtained for the illustrative value of
$p_t=80$ GeV and fixing jet rapidities at -2.5 and 0.9 units
respectively. We have performed the integration over the functions
$f_{ij}^{3/2}$ for the non-perturbative component (given in eq.
\eqref{eq:int}) numerically. We also assumed that the rapidity gap
has width $\Delta\eta=1$.

We notice that the effect of the term $D_{ij}^{\mathrm{NP}}$ is in
fact very small and can be safely ignored which is as an indication
that neglected higher orders in the expansion of $D_{ij}$ are even
more suppressed. We obtain as expected only a few percent correction
to the resummed perturbative result even at fairly low $E_t$, which
given the remaining theoretical uncertainties (notably having only
the large $N_c$ control over even leading logarithms) is not
significant in itself. However our main motivation as stressed in
the introduction, has not been the immediate phenomenology of the
$E_t$ flow observable, but rather a study of how power corrections
may be included in resummed predictions for general observables in
hadronic dijet production, since the soft function we treated here
generically appears in resummation of global hadronic dijet
observables.

In conclusion, we have carried out for the first time a calculation
of power corrections to transverse energy flow associated with dijet
production in hadron-hadron collisions. We illustrated the
computation for the energy flow distribution using the process
$q\bar{q}\rightarrow q\bar{q}$. The generalisation to other channels
is straightforward and only requires the numerical computation of
the diagonalised anomalous dimensions and the corresponding hard and
soft matrices. We found that the result does not correspond to the
usual shift found in studies of two-jet event shapes and energy
flows. The reason for this is the non-trivial colour algebra
involved in the case of hadron collisions. The techniques we used
here should enable better estimates of power corrections for
observables which have a similar nature to the one we introduced
here, such as the global inclusive jet cross-section we mentioned
earlier.

We re-emphasise that the simple calculation of this paper is just a
first step in the full quantitative estimate of power corrections,
pending the inclusion of two-loop effects. In the simpler cases of
$e^{+}e^{-} \to 2$ jets and DIS (1+1) jet processes, two-loop
effects merely rescaled the simple one gluon calculation by a
universal (Milan) factor \cite{Milan, universality, universality2,
universality3}, which arises when considering a two-loop analysis
for the argument of the coupling. We shall leave the inclusion of
such effects to forthcoming work.

\acknowledgments{We would like to thank Lorenzo Magnea, Giuseppe
Marchesini and Gavin Salam for useful discussions and Lorenzo Magnea
for comments on the manuscript.}

\appendix

\section{The functions $f_{ij}$}

We present here expressions for the functions $f_{ij}$. We define
$\eta_3$ and $\eta_4$ to be the rapidities of the outgoing hard
legs. We specify the kinematics of the particles as follows:
\begin{eqnarray}
p_1 &=& x_1\sqrt{s}/2 \left(1,0,0,-1\right),\\
p_2 &=& x_2\sqrt{s}/2 \left(1,0,0,1\right),\\
p_3 &=& P_t \left(\cosh\eta_3,1,0,\sinh\eta_3\right),\\
p_4 &=& P_t \left(\cosh\eta_4,-1,0,\sinh\eta_4\right),\\
k&=& k_t \left(\cosh\eta,\cos\phi,\sin\phi,\sinh\eta\right),
\end{eqnarray}
where $s$ is the hadronic centre-of-mass energy squared, related to
$\hat{s}$ (the partonic centre-of-mass energy squared) by
$\hat{s}=x_1x_2 s $, with $x_1$ and $x_2$ being the momentum
fractions of the incoming protons, carried by the incoming partons
``1'' and ``2'' respectively.

The functions $f_{ij}(\eta,\phi)=k_t^2\, p_i.p_j/(p_i.k \, p_j.k)$,
are given by:
\begin{eqnarray}
f_{12} & = & 2,\\
f_{13} & = & \frac{e^{\eta_3-\eta}}{\cosh(\eta_3-\eta)-\cos\phi},\\
f_{14} & = & \frac{e^{\eta_4-\eta}}{\cosh(\eta_4-\eta)+\cos\phi},\\
f_{23} & = & \frac{e^{-\eta_3+\eta}}{\cosh(\eta_3-\eta)-\cos\phi},\\
f_{24} & = & \frac{e^{-\eta_4+\eta}}{\cosh(\eta_4-\eta)+\cos\phi},\\
f_{34} & = & \frac{\cosh(\eta_3-\eta_4)+1}{(\cosh(\eta_3-\eta)-
\cos\phi)(\cosh(\eta_4-\eta)+\cos\phi)}.
\end{eqnarray}

\end{document}